# RETHINKING CASE FATALITY RATIOS FOR COVID-19 FROM A DATA-DRIVEN VIEWPOINT

Phoebus Rosakis[a,b,*], PhD & Maria E. Marketou[c], MD, PhD

[a]Department of Mathematics and Applied Mathematics, University of Crete, Heraklion 70013, Greece
[b]Institute of Applied and Computational Mathematics, Foundation of Research and Technology-Hellas, Heraklion 70013, Greece
[c]Department of Cardiology, Heraklion University Hospital, Heraklion 70013, Greece
[*]rosakis@uoc.gr

While examining the association between the Case Fatality Ratio (*CFR*) and cumulative number of COVID-19 infections in this journal, Kenyon[1] recently came across various difficulties in estimating the *CFR*. One of these was addressed by Baud and colleagues[2], who pointed out that the *CFR* (number of reported deaths divided by reported cases) ignores the time delay between incubation and death. Various problems[3] with their approach have been identified, but a concrete solution is unclear. While we agree that time lag plays an important role, it is overestimated by Baud and coworkers[2], whereas reported *CFR* values[4] ignore time lag completely. We find that either of these approaches introduces a spurious time dependence that severely distorts the magnitude and true meaning of the *CFR*. Instead, a suitably corrected *CFR* is far more useful as an indicator of COVID-19 fatality, because it turns out to be constant in time for many countries, as we show.

The *CFR* is unfavorably compared with the Infection Fatality Ratio (*IFR*)[2-6] of deaths over total actual infections, often because asymptomatic cases do not contribute to it, unless identified by testing. The *IFR* is important, but practically impossible to measure, due to lack of data for the denominator, which requires widespread, continuous random testing[7].

The *CFR* (only including reported cases) may have its uses in estimating fatalities[1]. Assuming random testing is very limited[4], the majority of reported cases have developed symptoms severe enough to seek medical assistance; these individuals are far more likely to die from the disease than asymptomatic cases[8], which would go undetected in the absence of testing. In this sense, the *CFR* is a meaningful measure of fatality risk among symptomatic individuals. This begs the question whether *CFR* versus time might be roughly constant for each country, at least during a period of fixed social distancing measures. This constant value would be different for each country, because of differing age distributions, mortality being a strongly increasing function of age[8], and possibly other factors.

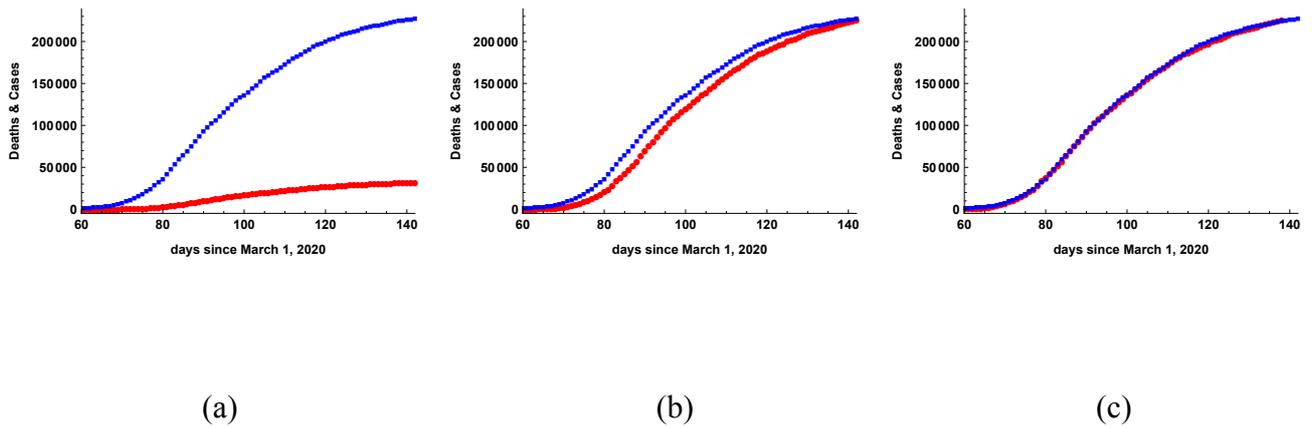

(a) (b) (c)

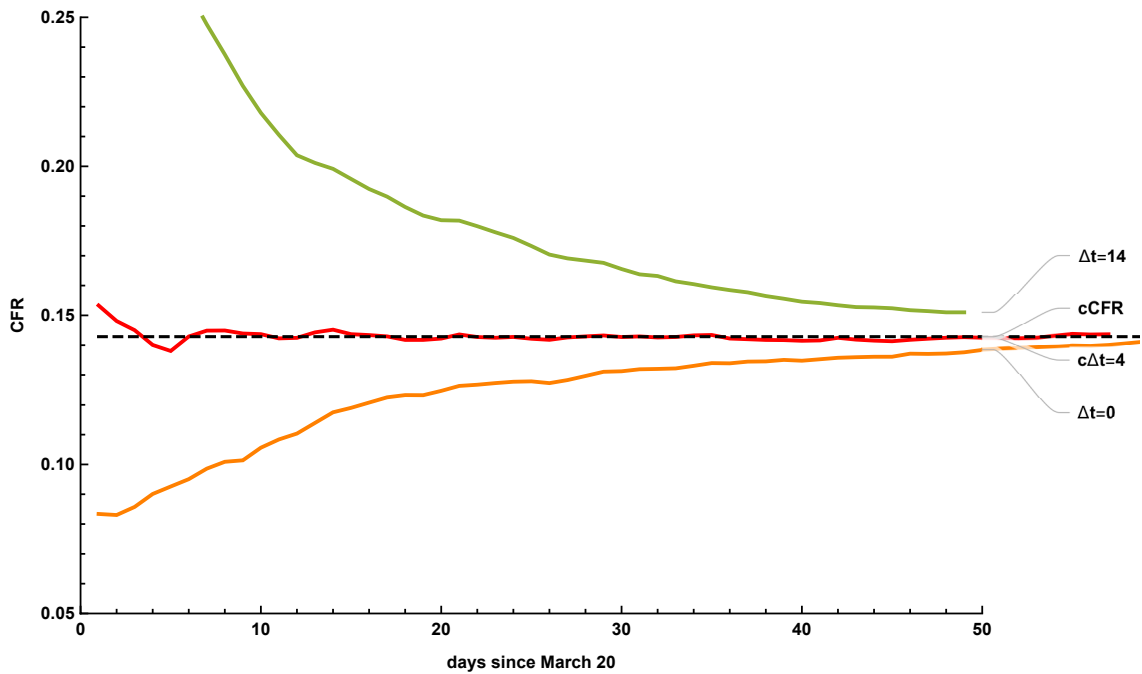

(d)

**Figure1: The case for a constant CFR.** (a) Italy deaths (red) and cases (blue). (b) as in (a) but with deaths scaled by a factor of 7. (c) as in (a) but with deaths scaled by 7 and shifted back by 4 days. The result is a constant CFR=1/7. (d) Various estimates of *CFR* versus time (March 20—May 20, 2020) for Italy. Orange: reported value[4] ignoring time delay (*Δt=0*). Black dashed line: our prediction of *cCFR=0·14*. Red: Corrected CFR with deaths shifted back by our predicted *cΔt=4* days. Green: Using Baud *et al.* method[1] with deaths shifted back by *Δt=14* days.

At first glance, this hypothesis is not supported by the COVID-19 data[4]. Most countries have an increasing, and some a decreasing *CFR* that eventually levels off to a constant.

To test the constant *CFR* hypothesis, we started with a hard-hit country, Italy, plotted deaths and reported cases versus time (figure 1a), and observed that multiplying deaths by roughly a factor of 7, made the two graphs almost the same (figure 1b), except for a shift *Δt=4* days; after compensating for which they became nearly indistinguishable (figure 1c), implying a *CFR*≈1/7=0·14 that remains virtually constant within 3% of 0·14.

Baud and coworkers[2] used a lag of 14 days, representing symptom onset to death. Instead we feel that the time lag *Δt* should reflect time from reporting to death. Delays from onset to reporting do occur[5,9]. In Singapore these delays had a mean of a week[10] and could exceed two weeks. Moreover, delays in reporting bring delays in critical medical care, hence may accelerate death. By increasing the time from onset to reporting, such factors decrease the time lag *Δt* from reporting to death, so we might expect *Δt* to be much less than 14 days, but uncertainty is introduced. Here, instead of arbitrarily picking *CFR* and *Δt*, or using estimates from a different location[5], we let the data decide. Data-driven predictions[11] of epidemic metrics are promising. We use a simple data-driven approach to find the right constant values, *cCFR* and *cΔt*. Simply put, we choose these values to be the ones that minimize the root mean square deviation between cases and deaths versus time, with deaths multiplied by *cCFR* and shifted back by *cΔt*. See the Appendix for details. This gives *cΔt=4* days for Italy. Shifting deaths back in time by this *cΔt*, then dividing by cases, yields a virtually constant *CFR* versus time (red curve, figure 1d), equal to *cCFR*≈0·14 (black dashed line, figure 1d) within a few percent. The statement *"14% of reported cases die after four days"* remains closer to the truth for much longer than any analogous statement regarding the reported, variable *CFR* that nearly doubles its value in two months (orange curve figure 1d).

This procedure works for many countries (figure 2), producing a different *cCFR* and *cΔt* for each, but also for the entire world: *cCFR*≈0·08, *cΔt=3* days (black dashed line, figure 2), but a nearly constant corrected CFR for all cases considered.

The reported[4] *CFR* (orange curve, figure 1d), which ignores time lag, increases with time and underestimates Italy's *cCFR* by a time-dependent amount. Baud et al. approach, shifting deaths back by *Δt*=14 days[2] (green curve, figure 1d) overestimates Italy's *cCFR* and decreases with time.

In summary, by allowing for an initially unknown time lag between case reporting and death, we find that many countries, and the entire world, exhibit a corrected *CFR* that is essentially constant during a long period of imposed social distancing. This value can be estimated long before the full evolution of the pandemic, hence it is useful for early prediction of fatalities, in situations where extensive random testing is not available.

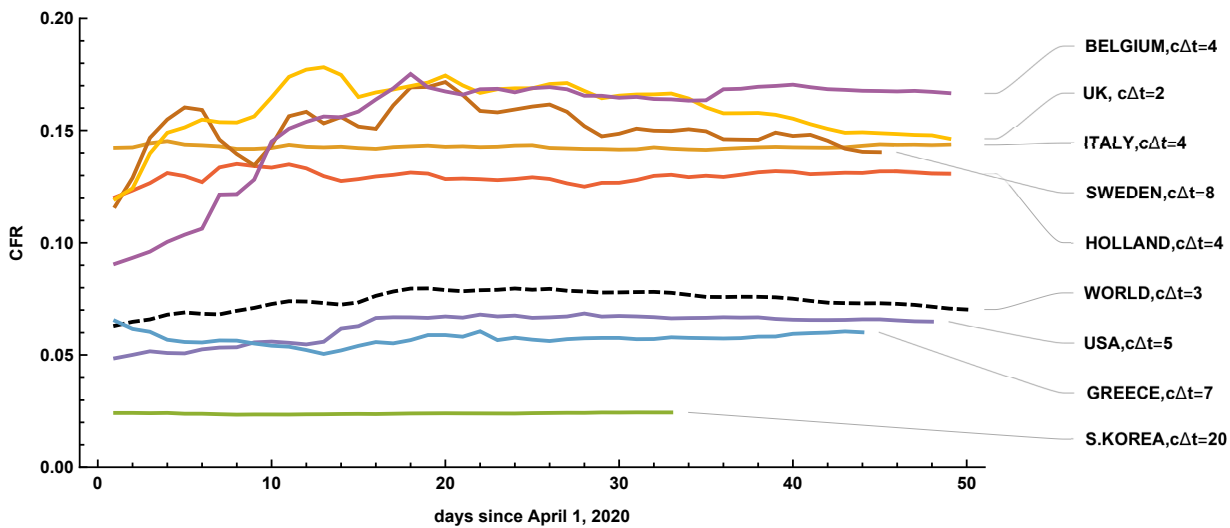

**Figure 2: Corrected Case Fatality Rate** versus time (April 1—May 20, 2020) for eight countries and the world, taking into account optimal time delay *cΔt* from reporting to death for each country. Our approach yields CFR versus time that is remarkably close to a constant for each country.

**Authors' contributions:** PR and MEM designed the research. PR performed the analysis with input from MEM. Both authors discussed the results and wrote the manuscript.

**Funding:** None.

**Ethical approval:** The work involved a secondary analysis of public access data. No ethics approval was necessary.

**Informed consent:** Not applicable.

**Declaration of Competing Interest:** The authors declare that they have no competing interests.

**Supplementary materials:** Details of our method can be found in the Appendix.

# SUPPLEMENTARY APPENDIX

We explain some details of our approach.

**Motivation.** Here we use Italy as an example. Suppose the number of total deaths reported by day $t$ is $D(t)$, and the number of total cases reported by day $t$ is $C(t)$. Then the usual definition of (reported) *CFR* is $CFR(t) = D(t)/C(t)$. Typically, this has the shape of the orange curve in figure 1d, and for Italy it roughly doubles its size from March 20 to May 20. On the other hand, we observed that for many countries, the two curves $D(t)$ and $C(t)$ versus $t$ (figure 1a) appeared similar, so multiplying deaths by a suitable constant factor might make them roughly equal to cas-es. When this is done it yields figure 1b where we plot *C(t)* and *7D(t)* together. The two curves appear to be shifted by about 4 days. Shifting the death curve back by 4 days gives *D(t+4)*. The shift reflects the typical value of time delay between reporting and death. Here it means that a case that dies at day *t+4* has been reported about 4 days earlier at day *t*. After we shift the rescaled death curve back by 4 days, namely *7D(t+4)*, it becomes almost indistinguishable with the reported case curve *C(t)*, figure 1c, so we have *C(t)≈7D(t+4)* or equivalently

$$D(t+4)/C(t) \approx 1/7 \approx 1.4$$

The left hand side of this for the Italy data[7] is nearly constant, and stays with 3% of 0·14 from April 20 onwards.

This means an essentially constant corrected *cCFR*~0·14, corresponding to a time delay *Δt=4* days. Of course here the factor of 7 and the time delay of 4 days were chosen by trial and error, so we next describe a more objective, unbiased method of obtaining them.

**Algorithm for Finding *cCFR* AND *cΔt*.** Rescale the death curve by a constant factor $K$ and shift it backwards in time by an integer constant $z$ (days) to obtain the graph of the function $KD(t+z)$. We provide a systematic way of automatically finding the values of $K$ and $z$ that bring the two curves ( *C(t) versus t* and $KD(t+z)$ *versus t*) as close as possible, without recourse to trial and error. We recall that a measure of their difference is the root-mean-square deviation (RMSD).This is the square root, of

the mean, of the squared difference between the values of *C(t)* and *KD(t+z)*, over all times considered:

$$RMSD(K,z) = \sqrt{\sum_{t=1}^{N} \frac{1}{N}[C(t) - KD(t+z)]]^2}$$

The sum is from day 1 to day N. Here we typically choose day 1 as March 20, 2020, by which time government restrictions were in place in most countries[7] and the pandemic has evolved past the initial stage (of very small case and death numbers). Day N is May 20, 2020 (final date of data used).

Since *C(t)* and *D(t)* are known functions from data[7], *RMSD* reduces to a function of *K* and *z*, and we find its minimum. We let *cCFR* and *cΔt* be the values of *K* and *z* respectively that minimize the RMS deviation. We then plot the quantity $D(t + c\Delta t)/C(t)$ versus t, (red curve, figure 1d) We observe that this quantity is much closer to the constant value *cCFR* than the reported ratio $CFR(t) = D(t)/C(t)$ (orange curve, figure 1d). For Italy, it remains within 3% of *cCFR*, for the world within 10% of *cCFR=0·08* after April 1, 2020, whereas the reported ratio $CFR(t) = D(t)/C(t)$ for Italy increases from 0·08 to 0·014 during the same time.

We have tested this approach for 8 countries (color curves figure 2), and the entire world (dashed balck line, figure 2). The CFR for each of these is close to a constant in the time period shown (April and May) after the epidemic developed beyond the early stage, and after government-imposed restrictions were in place by mid-March[7] in most countries. At the early stages when cases and deaths are very few, the CFR can exhibit oscillations and artificial spikes due to back-shifting leading to a small divisor problem.

**Significance of *cΔt*.** Here *cΔt* is the average time delay between case reports and deaths. This equals the difference between the onset-to-death delay and the reporting delay (onset to reporting). Since reporting delays[9,10] can exceed a week, this explains why typical values are less than the 14 day onset-to-death period assumed by Baud and coworkers[1]. In the case of South Korea the large *cΔt*=20 days seems to be due to the extensive random testing that reported many cases long before they developed noticeable symptoms. In that case *cΔt* can be larger than the lag between onset of symptoms and death.